\documentclass[aip,jap,reprint,nofootinbib]{revtex4-1}
\usepackage{graphicx}% Include figure files
\usepackage{times}
\usepackage{mathptmx}
\usepackage{mathrsfs}
\usepackage{amsmath,amssymb}

\allowdisplaybreaks
\begin{document}
\abovedisplayskip=3pt plus 3.0pt minus 3.0pt
\abovedisplayshortskip=0.0pt plus 3.0pt
\belowdisplayskip=3pt plus 3.0pt minus 3.0pt
\belowdisplayshortskip=3pt plus 3pt minus 3pt

\title{\large Ultralow-frequency spontaneous oscillations of the current in polycrystalline silicon}
\author{\vspace{6pt}V.A. Dorosinets}
\author{N.A. Poklonski}
%\email{poklonski@bsu.by}
%\thanks{}
\author{V.A. Samuilov}
\author{V.F. Stel'makh}
\affiliation{\vspace{6pt}Belarusian State University, pr. Nezavisimosti 4, 220030 Minsk, Belarus; e-mail: poklonski@bsu.by}


\date{\today}

\begin{abstract}
\parbox{138mm}{\vspace{6pt}\raggedright
We report results of experimental study of ultralow-frequency spontaneous oscillations of the current in polycrystalline silicon films subjected to strong electric fields at room temperature.
}
\end{abstract}

%Use showkeys class option if keyword display desired

\maketitle %\maketitle must follow title, authors, abstract and \pacs


Instabilities resulting in periodic\cite{1,2,3} and stochastic (with a small number of degrees of freedom)\cite{2,3,4} spontaneous oscillations of the electric current in semiconductors are currently attracting much interest. We shall report the results of an experimental study of such oscillations observed for the first time in polycrystalline silicon films subjected to strong electric fields.

Polycrystalline silicon films of thickness 0.4~$\mu$m were grown by pyrolytic dissociation of monosilane at a low pressure on oxidized single-crystal silicon substrates. These films were doped by implantation of boron ions of 30~keV energy in doses of $1.3{\cdot}10^{12}$--$1.9{\cdot}10^{15}$~cm$^{-2}$. Annealing at 1000\,$^\circ$C for 30~min ensured a uniform distribution of the impurity across the film thickness and then photolithography was used to form resistive structures with aluminum contacts deposited on contact $p^+$-type regions which were alloyed. The average grain size deduced from an electron-microscopic examination of the replicas amounts to $\sim 110$~nm.

\begin{figure}%[!h]
\noindent\hfil\includegraphics{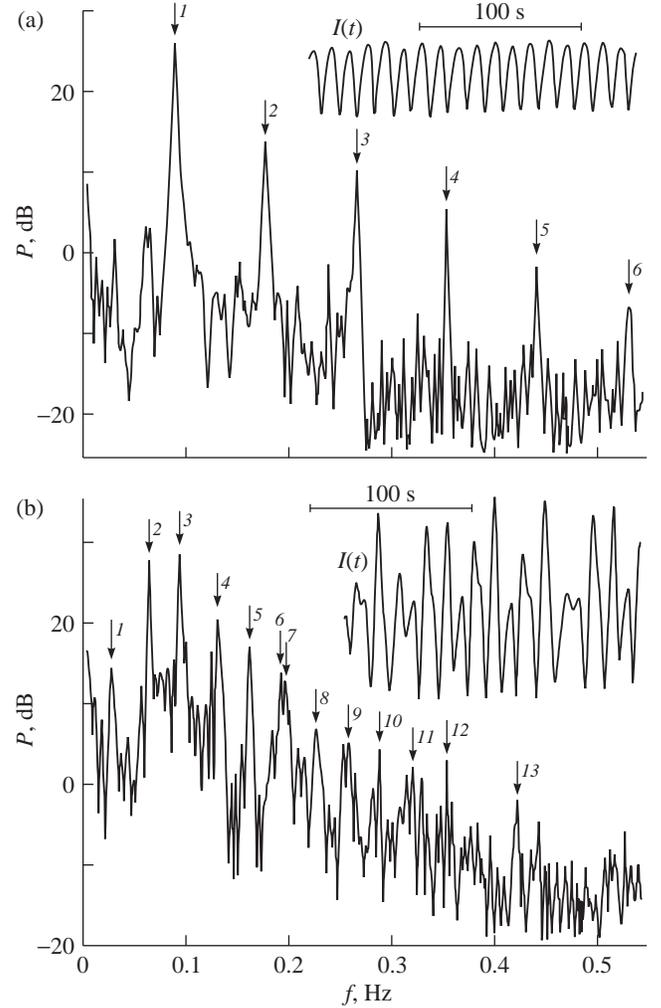}
\caption{Spectra of the signal power $P(f)$ and time dependences of the current $I(t)$ obtained at two fixed values of the electric field intensity: a) $E_1$; b) $E_2$, where $E_2 > E_1$. Arrows mark peaks: a) peak \emph{1} at $f_1$; \emph{2}~at $2f_1$; \emph{3} at $3f_1$; \emph{4} at $4f_1$; \emph{5} at $5f_1$; \emph{6} at $6f_1$; b) peak \emph{1} at $f_1 - f_2$; \emph{2}~at $f_2$; \emph{3} at $f_l$; \emph{4} at $2f_2$; \emph{5} at $f_1 + f_2$; \emph{6} at $2f_1$; \emph{7} at $3f_2$; \emph{8} at $2f_2 + f_1$; \emph{9} at $2f_1 + f_2$; \emph{10} at $3f_1$; \emph{11} at $2f_1 + 2f_2$; \emph{12} at $3f_1 + f_2$; \emph{13} at $3f_1 + 2f_2$}\label{fig:01}
\vspace{-3mm}
\end{figure}

The current-voltage characteristics of polycrystalline silicon (polysilicon) films were linear in fields $E < 10^3$~V/cm but in the range $E \gtrsim 10^3$~V/cm the current rose exponentially on increase in the voltage. In films with the boron concentration $N \lesssim 2{\cdot}10^{17}$~cm$^{-3}$ when the current-voltage characteristics were recorded under static conditions in strong electric fields ($E \gtrsim 10^4$~V/cm) it was found that undamped oscillations of the electric current of frequency from $5{\cdot}10^{-3}$ to $2{\cdot}10^{-1}$~Hz were observed at $T = 300$~K. The oscillation amplitude reached a few tens of percent of the steady-state value~\cite{Poklonskij85VKFP133}. The time dependence of the current $I(t)$ was subjected to the fast Fourier transform and this gave the power spectrum of the signal $P(f)$ (Fig.~\ref{fig:01}). We observed both periodic oscillations of the current (Fig.~\ref{fig:01}a) as well as quasiperiodic oscillations (Fig.~\ref{fig:01}b). In the case of periodic oscillations the current power spectrum consisted of the main part and its harmonics. In addition to recording $I(t)$, we also measured the signals from side potential contacts separated along a sample (these signals were applied to the $X$ and $Y$ inputs of a potentiometer via electrometric amplifiers). This produced a closed figure of near-elliptic shape, providing a direct proof of the motion of charge inhomogeneities\cite{5} in a film subjected to an external electric field. An increase in the field intensity resulted in quasistochastic behavior, manifested by a distortion of the time dependence of the current and also by an increase in the fundamental frequency as well as by the appearance of an additional frequency and linear combinations of frequencies (Fig.~\ref{fig:01}b).

\begin{figure}%[!h]
\noindent\hfil\includegraphics{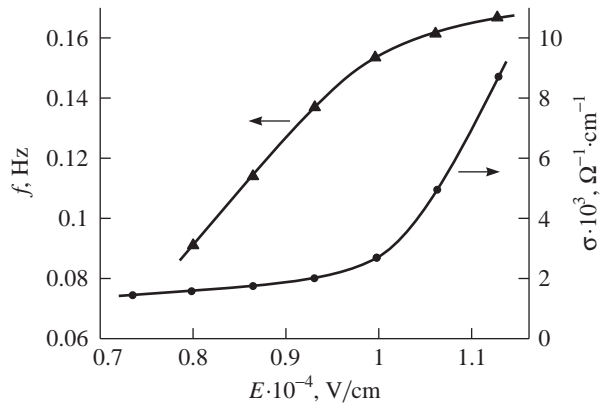}~~~~
\caption{Dependences of the frequency $f$ of the current and of the differential conductivity $\sigma$ on the electric field $E$}\label{fig:02}
\end{figure}

Such generation of low-frequency oscillations of the current had been observed earlier in experiments on compensated Ge single crystals and interpreted using a model of spatial trap-charging waves.\cite{5,6} However, an additional experimental investigation of the dependences of the oscillation frequency and of the electrical conductivity $\sigma$ on the voltage applied to a polysilicon sample (Fig.~\ref{fig:02}) and on the illumination of a sample, together with an analysis of the nature of the current-voltage characteristics revealed wave-like charge motion processes which were incompatible with this model. For example, the frequency of oscillations of the current in our samples increased on increase in the applied voltage and fell on increase in the illumination intensity. In the case of spatial trap-charging waves in compensated semiconductors the dependences should be exactly the reverse.

The mechanism of generation of oscillations of the electric current in polysilicon films should be as follows. It is known that the main parameters describing electrophysical properties of polysilicon films are the average grain size $L$, the density of trap states $Q_\text{t}$ at grain boundaries, and the dopant concentration $N$. In the case of moderate dopant concentrations ($N \simeq Q_\text{t}/L$) an increase in $N$ within one order of magnitude reduces the electrical resistivity $\rho$ by 5 to 6 orders of magnitude, whereas the Hall mobility $\mu_\text{H}$ is a nonmonotonic function of the dopant concentraton.\cite{7} Such dependences of $\rho$ and $\mu_\text{H}$ on $N$ can be explained by an increase in the height of the electric potential barriers at grain boundaries if $N < Q_\text{t}/L$ or a reduction in its barriers if $N > Q_\text{t}/L$. On the other hand, the height of these barriers in polysilicon can be controlled not only by varying the dopant concentration, but also by generation of carriers in the bulk and at grain boundaries as a result of illumination or injection from the contacts\cite{8} which may be manifested by a significant change in the electrical conductivity of a sample. For example, experimental studies of polysilicon films have revealed switching from a high- to a low-resistivity state\cite{9,10} explained by a reduction in the height of electric potential barriers for the majority carriers throughout the sample when barriers break down\cite{9} or because carriers are injected from the contacts.\cite{10} If the depth of injection is considerably less than the length of the sample, then for a specific relationship between the parameters (such as the voltage, temperature, dopant concentration, and grain size) we may observe instabilities, exactly as in the case of single-crystal samples.\cite{11} In our case a slow charge exchange between traps were observed and it varied along the samples. However, since traps in polysilicon were concentrated at grain boundaries, the dispersion law for the spatial barrier-charging waves observed by us differed from the dispersion law for spatial trap-charging waves in single-crystal compensated semiconductors.

%\newpage
%~\\
%\newpage
%\bibliography{poklonski}

\begin{thebibliography}{00}%
\bibitem{1}
L.L. Golik, V.E. Pakseev, Yu.I. Balkarei, M.I. Elinson, Yu.A. Rzhanov, and V.K. Yakushin, Fiz. Tekh. Poluprovodn. \textbf{18} (3), 502--507 (1984) [Sov. Phys. Semicond. \textbf{18}, 310 (1984)].
\bibitem{2}
L.L. Golik, V.E. Pakseev, M.I. Elinson, and V.K. Yakushin, Fiz. Tekh. Poluprovodn. \textbf{20} (11), 2084--2091 (1986) [Sov. Phys. Semicond. \textbf{20}, 1303 (1986)].
\bibitem{3}
S.V. Bumyalene, K.A. Piragas, Yu.K. Pozhela, and A.V. Tamashyavichyus, Fiz. Tekh. Poluprovodn. \textbf{20} (7), 1190--1194 (1986) [Sov. Phys. Semicond. \textbf{20}, 752 (1986)].
\bibitem{4}
K. Piragas, Yu. Pozhela, A. Tamashyavichyus, and Yu. Ul'bikas, Fiz. Tekh. Poluprovodn. \textbf{21} (3), 545--548 (1987) [Sov. Phys. Semicond. \textbf{21}, 335 (1987)].
\bibitem{Poklonskij85VKFP133}
N.A. Poklonskii, V.A. Samuilov, V.F. Stel'makh. \emph{Autowave processes in polycrystalline silicon films}. in Proc. of X all-USSR Conf. on Semiconductor Physics, Minsk 17--19 Sept. 1985, pp. 133--134. [in Russian].
\bibitem{5}
N.G. Zhdanova and M.S. Kagan, Fiz. Tekh. Poluprovodn. \textbf{15} (1), 168--170 (1981) [Sov. Phys. Semicond. \textbf{15}, 99 (1981)].
\bibitem{6}
N.G. Zhdanova, M.S. Kagan, and S.G. Kalashnikov, Fiz. Tekh. Poluprovodn. \textbf{17} (10), 1852--1854 (1983) [Sov. Phys. Semicond. \textbf{17}, 1182 (1983)].
\bibitem{7}
J.Y.W. Seto, J. Appl. Phys. \textbf{46} (12), 5247--5254 (1975).
\bibitem{8}
P.T. Landsberg and M.S. Abrahams, J. Appl. Phys. \textbf{55} (12), 4284--4293 (1984).
\bibitem{9}
B.A. Gezalov, F.D. Kasimov, V.A. Vetkhov, and V.M. Mamikonova, Pis'ma Zh. Tekh. Fiz. \textbf{9} (24), 1523--1526 (1983) [Sov. Tech. Phys. Lett. \textbf{9}, 652 (1983)].
\bibitem{10}
C.Y. Lu, N.C.C. Lu, and C.C. Shin, J. Electrochem. Soc. \textbf{132} (5), 1193--1196 (1985).
\bibitem{11}
Yu.K. Pozhela, Plasma and Current Instabilities in Semiconductors (Moscow, 1977) 368 p. [in Russian].
\end{thebibliography}

%
\end{document}